\documentclass[twocolumn,showpacs,preprintnumbers,amsmath,amssymb]{revtex4}

\usepackage{graphicx}
\usepackage{dcolumn}
\usepackage{bm}

\begin{document}

\preprint{APS/123-QED}

\title{Nonequilibrium evolution thermodynamics}

\author{Leonid S. Metlov}

 \email{lsmet@kinetic.ac.donetsk.ua}
\affiliation{Donetsk Institute of Physics and Engineering, Ukrainian
Academy of Sciences,
\\83114, R.Luxemburg str. 72, Donetsk, Ukraine
}%

\date{\today}

\begin{abstract}
A new approach - nonequilibrium evolution thermodynamics, is compared with classical variant of Landau approach.
\end{abstract}

\pacs{05.70.Ln; 05.45.Pq}
\maketitle

\section{Introduction. An example from theory of ideal gases}

Well known that the internal energy $U$ is function of entropy $S$ and volume $V$ of a system: 
  \begin{equation}\label{b1}
U=U(S,V).
  \end{equation}

So well known that the free energy $F$ is function of temperature $T$ and volume $V$ of a system:
  \begin{equation}\label{b2}
F=F(T,V).
  \end{equation}

We can say, that entropy and volume are eigen-arguments of the internal energy, so as temperature and volume are eigen-arguments of free energy. Well known also that internal energy for ideal gases is simply expressed versus temperature \cite{b64}:
  \begin{equation}\label{b3}
U=U_{0}+C_{V}T,
  \end{equation}
where $C_{V}$ is heat capacity at constant volume.

In the same time, the internal energy versus eigen-arguments for an ideal gas looks as a complex thing. Really, the expression for entropy of ideal gases is
  \begin{equation}\label{b4}
\frac{S}{\nu}-S_{0}=C_{V}\ln T+R \ln \frac{V}{N}=\ln[(T)^{C_{V}}(\frac{V}{N})^{R}],
  \end{equation}
where $\nu$ is number of moles, $S_{0}$ is some constant, $R$ is molar gas constant, $N$ is number of particles \cite{b64}. From (\ref{b4}) one can give expression for temperature and for internal energy:
\begin{eqnarray}\label{b5}
\nonumber
T=(\frac{V}{N})^{\frac{R}{C_{V}}}\exp(\frac{S}{\nu}-S_{0}),  \\
U=C_{V}(\frac{V}{N})^{\frac{R}{C_{V}}}\exp(\frac{S}{\nu}-S_{0}),
\end{eqnarray}
Comparing with (\ref{b3}) one can see that the internal energy versus its eigen-arguments looks more complicated than versus a foreign variable. Nevertheless, the expression for the internal energy in the form (\ref{b5}) is more correct from the point of view of thermodynamics. At least one can calculate the temperature $T$ and pressure $P$ with help of standard procedure by means of differentiation of the internal energy:
  \begin{equation}\label{b6}
T=\frac{\partial U}{\partial S},  
\quad P=-\frac{\partial U}{\partial V},
  \end{equation}

One can easily prove the identity of temperature following from Eqs. (\ref{b5}) and (\ref{b6}) with help of equation of state for ideal gases $RT=PV$.

Thus, in this example we see, that some relations of thermodynamics look simpler versus variables, which isn't its eigen-arguments. In the next issues we will see that this 'rule' is executed and in more general cases.

\section{Principle of minimum of free energy}

If a solid consisting $N$ particles has $n$ structural defects (e.g., vacancies, substituted atom, ets.) then the equilibrium or steady state in this case can be found from the maximum of probability distribution function taking in the form \cite{f55, ll69, s89, g97}:
  \begin{equation}\label{b7}
f(n)=C\dfrac{(N+n)!}{N!n!}\exp(-\dfrac{U(n)}{kT}),
  \end{equation}
where $C$ is a normalized factor, $U(n)$ is the internal energy depending of number of structural defects, $k$ is Boltzmann's constant. The pre-exponential factor describes the combinational, that is, entropic, part of the distribution function, connected with degeneration of microstates. The exponential factor describes the restrictive part of the distribution function, connected with the overcoming of potential barriers between microstates. In a quadratic approximation
  \begin{equation}\label{b8}
U=U_{0}+u_{0}n-\frac{1}{2}u_{1}n^2.
  \end{equation}
where $u_{0}$ and $u_{1}$ are some constants.

Bringing variables independing of number of defect $n$ into the inessential constant $C$ the expression (\ref{b7}) can be written down in the form:
  \begin{equation}\label{b9}
f(n)=C\dfrac{(N+n)!}{n!}\exp(-\dfrac{u_{0}n-\frac{1}{2}u_{1}n^2}{kT}),
  \end{equation}
or in a form of product:
  \begin{equation}\label{b10}
f(n)=C\prod_{l=1}^{N}(n+i)\exp(-\dfrac{u_{0}n-\frac{1}{2}u_{1}n^2}{kT}).
  \end{equation}

By differentiating it, we obtain
  \begin{equation}\label{b11}
\frac{\partial f(n)}{\partial n}=(\sum_{k=1}^{N+n}\frac{1}{k}-\sum_{k=1}^{n}\frac{1}{k}-\frac{u_{0}-u_{1}n}{kT})f(n).
  \end{equation}

The extreme meaning of probability distribution function is at n, which obeys next transcendental equation:
  \begin{equation}\label{b12}
\sum_{k=1}^{N+n}\frac{1}{k}-\sum_{k=1}^{n}\frac{1}{k}-\frac{u_{0}-u_{1}n}{kT}=0.
  \end{equation}

From a table value partial sums one can find \cite{gr07}:
  \begin{equation}\label{b13}
\sum_{k=1}^{n}\frac{1}{k}=C+\ln n +\frac{1}{2n},
  \end{equation}
where $C$ is some constant. By substituting (\ref{b13}) into (\ref{b12}), for case $N>>n>>1$ we obtain:
  \begin{equation}\label{b14}
n=N\exp(-\frac{u}{kT}),
  \end{equation}
where
  \begin{equation}\label{b15}
u\equiv\frac{\partial U}{\partial n}=u_{0}-u_{1}n
  \end{equation}
is energy of a defect. As is evident from the last formula the energy of defect is not strictly constant, but depends from total number of defects. The relation (\ref{b14}) is equation of state for an equilibrium case, the relation (\ref{b15}) is equation of state too, but for more general nonequilibrium case included the equilibrium state as a partial case. It is need to consider the Eqs. (\ref{b14}) and (\ref{b15}) together, as a set of equations for deducing both the energy of defect $u_{e}$ and density of defects $n_{e}$ into the equilibrium state. 

Thus, the equation of state (\ref{b14}) is obtained from the condition of most probability state as maximum of probability distribution function (\ref{b7}). Same result one can obtain from the principle of minimum of the free energy. Really, pre-exponential factor in (\ref{b7}) is the thermodynamic probability \cite{f55, bbe92}
  \begin{equation}\label{b16}
W=\dfrac{(N+n)!}{N!n!},
  \end{equation}
the logarithm of which is configurational entropy $S_{c}=k\ln W$. Note, that configurational entropy is one-valued function of number of defects. It is perfectly independent of energy of defect (and of temperature too). 

Then described above procedure can be schematically displayed as \cite{s89}
  \begin{equation}\label{b17}
W\exp(-\dfrac{U(n)}{kT})\rightarrow max
  \end{equation}
or after logarithmic operation in the form
  \begin{equation}\label{b18}
\ln W-\dfrac{U(n)}{kT}\rightarrow max.
  \end{equation}
Inverting the sign, we come to the free energy minimization principle
  \begin{equation}\label{b19}
U(n)-kT\ln W\equiv U-TS_{c}=F_{c}\rightarrow min.
  \end{equation}

Nevertheless, this excellent result contains a contradiction. Really, the product $TS_{c}$ entered into definition of the free energy $F_{c}$ is bounded energy, which is lost for a production of the work by a system. In another side, the total energy of defects in the main part is physically energy, which is lost for the production of the work too. Only a little part of it remains for the work production. Then we can write down that
  \begin{equation}\label{b20}
TS_{c}\approx un
  \end{equation}

And now we can introduce a new specific kind of free energy by means of subtraction of bounded energy in the form product $un$ from internal energy (\ref{b8}). 
  \begin{equation}\label{b21}
\tilde{F_{c}}=U-un=U_{0}+\dfrac{1}{2u_{1}}(u_{0}-u_{V})^{2}.
  \end{equation}
Here we use equation of state (\ref{b15}) for elimination of density of defects. It is very easy to establish that
  \begin{equation}\label{b22}
n=-\frac{\partial \tilde{F_{c}}}{\partial u}.
  \end{equation}
Both relations (\ref{b15}) and (\ref{b22}) are connected couple of equations between the internal energy $U$ and the modified configurational free energy $\tilde{F_{c}}$ from one side, and between density of defects $n$ and energy of defect $u$ from another side. One can see that the energy of defect is eigen-argument for the internal energy, and density of defects is eigen-argument for the modified configurational free energy. In this connection, the exact free energy $F_{c}$ according to (\ref{b16}) and (\ref{b19}) is expressed through variable $n$, which isn't its eigen-argument.

We have same situation as in the previous section example for ideal gases. Namely, the free energy expressing versus the foreign «argument» obeys simple fundamental feature: minimization principle, as it expressing versus the eigen-argument don't obey this feature, and we must use additional operations for finding of equilibrium parameters. But from the thermodynamic point of view, the expression of free energy versus eigen-argument is more correct, as allows to use notations closed to the equilibrium thermodynamics in nonequilibrium cases. 

\section{KINETIC EQUATIONS}

\begin{figure*}
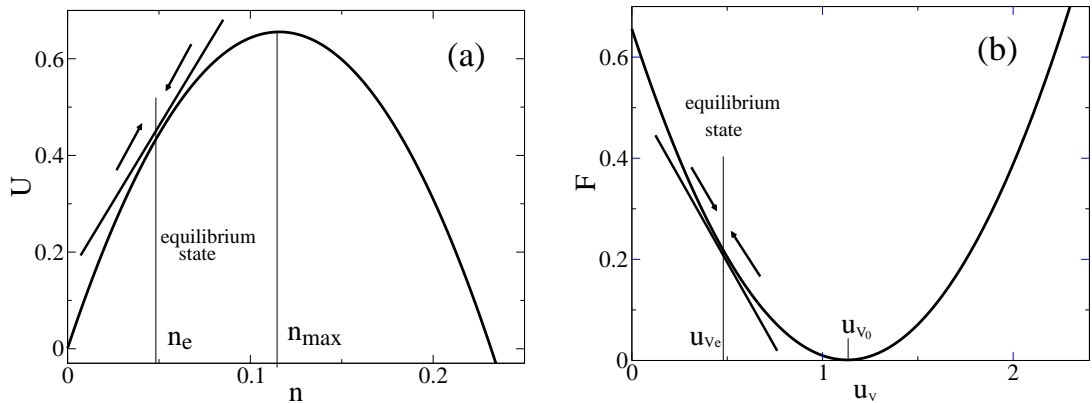

\includegraphics [width=2.7 in]{fig_1a}
\quad \quad
\includegraphics [width=2.7 in]{fig_1b}
\caption{\label{f1} Plots of the internal (a) and free (b) energy versus its eigen arguments. Tendency of the system to the equilibrium or steady state is indicated by arrows.}
\end{figure*}

Because the energy, needed for the formation of a new defect, is smaller in the presence of others than in defect-free crystal, the quadratic term in (\ref{b8}) has negative sign. Note that expression (\ref{b8}) is true both for equilibrium and non-equilibrium states. In this approximation the internal energy is a convex function of the defect number having the maximum at point $n=n_{max}$, as it is shown in Fig. \ref{f1} a.
 In same approximation the modified configurational free energy is a concave function with the minimum at point $u_{V}=u_{V0}$, as it is shown in Fig. \ref{f1}b.

With relationships (\ref{b14}) and (\ref{b15}), it is easy to show that the steady state corresponds neither to the maximum of the internal energy nor to the minimum of the free energy. The steady state is at point $n=n_{e}$, where 
  \begin{equation}\label{b23}
u_{e}=\dfrac{\partial U}{\partial n_{e}},  
\quad n_{e}=-\dfrac{\partial \tilde{F_{c}}}{\partial u_{e}}.
  \end{equation}
Here the additional subscript $e$ denotes the equilibrium value of a variable.

If a system has deviated from the steady state, then it should tend back that state with a speed, which is higher, the larger is the deviation \cite{m07, m08, m09}:
  \begin{equation}\label{b24}
\dfrac{\partial n}{\partial t}=\gamma_{n}(\dfrac{\partial U}{\partial n}-u_{e}),  
\quad \dfrac{\partial u}{\partial t}=-\gamma_{u}(\dfrac{\partial \tilde{F_{c}}}{\partial u}-n_{e}),
  \end{equation}

Both variants of the kinetic equations are equivalent and their application is a matter of convenience. The form of kinetic equations (\ref{b24}) is symmetric with respect to the use of internal and configurational free energy. In the right-hand parts of Eq. (\ref{b24}) the signs are chosen, based on solution stability, so that the internal energy is a convex function, and the free energy is a concave one. In the right side of the well-known Landau-Khalatnikov kinetic equation \cite{lh54}
  \begin{equation}\label{b25}
\dfrac{\partial n}{\partial t}=-\gamma \dfrac{\partial F_{c}}{\partial n}
  \end{equation}
the ``chemical potential'' is used in the form:
  \begin{equation}\label{b26}
\mu =\dfrac{\partial F_{c}}{\partial n}.
  \end{equation}

From the thermodynamic point view, such kind of variable isn't chemical potential in really, as it is specified by foreign corresponding to the free energy 'argument'. But it does not hind using this notation in practical work, as it directly realizes the minimization principle for the free energy.

If we consider that equilibrium energy of defect $u_{e}$ and number of defects $n_{e}$ slowly change during external action then we can introduce them under differentiation sign in (\ref{b24}) and definite new kind (shifted) of internal and free energy.
  \begin{equation}\label{b27}
\bold{U} =U-u_{e}n,  
\quad \tilde{\bold{F_{c}}} =\tilde{F_{c}}-un_{e}.
  \end{equation}

Then equations (\ref{b24}) are simplified a little:
  \begin{equation}\label{b28}
\dfrac{\partial n}{\partial t}=\gamma_{n}\dfrac{\partial \bold{U}}{\partial n},  
\quad \dfrac{\partial u}{\partial t}=-\gamma_{u}\dfrac{\partial \tilde{\bold{F_{c}}}}{\partial u}.
  \end{equation}

The original potentials $U$ and $\tilde{F_{c}}$ are connected by means of a Legendre-like transformation:
  \begin{equation}\label{b29}
F_{c}=U-un.
  \end{equation}

The shifted potential $\bold{U}$ and $\tilde{\bold{F_{c}}}$ are connected by means of transformation:
  \begin{equation}\label{b30}
\tilde{\bold{F_{c}}}=\bold{U}-un+un_{e}-u_{e}n,
  \end{equation}
which differs from the Legendre-like transformation by Poisson-like bracket $[un]=un_{e}-u_{e}n$.

The stationary point for the shifted potentials is coincided with a maximum of $\bold{U}$ and with a minimum of $\tilde{\bold{F_{c}}}$. Thus $\bold{U}$ is some effective thermodynamic potential, for which tendency of the original part of internal energy to minimum is completely compensated by entropic factor. Twice modified configurational free energy $\tilde{\bold{F_{c}}}$ tends to minimum, but this tendency is differ from it for the original configurational free energy $F_{c}$. The effective thermodynamic potential $\tilde{\bold{F_{c}}}$ tends to minimum in space of eigen-argument $u$, then the original free energy $F_{c}$ tends to minimum in the space of foreign 'argument' $n$.

\section{CONCLUDING REMARKS}

In the paper, a phenomenological approach, based on generalization of Landau technique is considered. For fast processes thermal fluctuations have no time to exert essential influence and it is possible to consider the problem in the mean-field approximation. The approach is based not on an abstract order parameter but on a physical parameters of structural defects -- their quantity (density) and the average energy. The new more general form of kinetic equations, symmetric with respect to using the internal energy $U$ and the modified configurational free energy $\tilde{F_{c}}$, is proposed. In this case, the density of defects and defect energy are related by a symmetric differential dependences of type (\ref{b15}), (\ref{b22}) and (\ref{b23}). Because the defect energy in the steady state is not equal to zero, the extreme principle of equality to zero of the derivative of free energy with respect to 'order parameter' in the framework of nonequilibrium evolution thermodynamics breaks down. This principle needs to be substituted with principle of the tendency to a steady state. Steady-state characteristics can not be determined in the framework of phenomenological approach, statistical and microscopic approaches are required.

The present form of kinetic equations can be generalized to all types of regularly or randomly distributed defects.

\begin{acknowledgments}
The work was supported by the budget topic № 0106U006931 of NAS of Ukraine and partially by the Ukrainian state fund of fundamental researches (grants F28.7/060). The author thanks A. Filippov for helpful discussions. The author thanks also him Referee for useful remarks and comments. 
\end{acknowledgments}

\end{document}